\documentclass{ifacconf_amx}

\usepackage{graphicx}      
\usepackage{amsmath,amssymb,amsfonts} 
\usepackage{xcolor}        
\usepackage{cite}          

\usepackage{siunitx}       
\usepackage{fancyhdr}      

\usepackage{lipsum}        

\usepackage[scaled]{helvet}
\usepackage[T1]{fontenc}

\newcommand{\ISBN}{UNKNOWN ISBN}

\newcommand{\PaperTitle}{A title for your paper}
\newcommand{\PaperDate}{01 April 1900}
\newcommand{\AuthorFooter}{Mustermann, M.,Supervisor, S., Oksanen T.}

\makeatletter
\let\old@ssect\@ssect 
\makeatother

\usepackage[hidelinks]{hyperref}      
\usepackage{cleveref}      

\makeatletter
\def\@ssect#1#2#3#4#5#6{%
	\NR@gettitle{#6}
	\old@ssect{#1}{#2}{#3}{#4}{#5}{#6}
}
\makeatother
\renewcommand{\ISBN}{978-3-911430-10-4}
\renewcommand{\PaperTitle}{Smart Prism with Tilt Compensation for CAN bus on Mobile Machinery Using Robotic Total Stations}
\renewcommand{\PaperDate}{28 February 2026}
\renewcommand{\AuthorFooter}{Sharma, S., Moll, M. and Oksanen, T.}

\usepackage{tabularx}
\usepackage{siunitx}
\usepackage[commandnameprefix=ifneeded,commentmarkup=footnote]{changes} 
\begin{document}
\begin{frontmatter}

   \title{\PaperTitle}

   \author[First]{Sumesh Sharma}
   \author[First]{Marcel Moll}
   \author[First]{Timo Oksanen}

   \address[First]{Technical University of Munich, Germany; Professorship of Agrimechatronics; e-mail: first.last@tum.de; \\ Munich Institute of Robotics and Machine Intelligence (MIRMI)}

   \begin{abstract}
Accurate reference trajectories are required to validate autonomous agricultural robots and highly automated off-road vehicles under real-world field conditions. In practice, robotic total stations provide millimeter-level prism center coordinates, but the point of interest on the vehicle is typically displaced by a lever arm, ranging from decimeters to multiple meters. Roll and pitch motions, as typically observed in off-road machinery, therefore introduce horizontal point of interest errors far exceeding the measurement accuracy of robotic total stations observations. This paper presents the design, implementation, and validation of a Smart Prism prototype that augments a robotic total station prism with an inertial measurement unit to enable real-time tilt compensation. The prototype integrates an STM32H7 microcontroller and a Murata SCH16T-series IMU and estimates roll and pitch angles using an adaptive complementary filter. The tilt-compensated point of interest coordinates are obtained by transforming a calibrated lever arm from the body frame into the navigation frame and combining it with robotic total station prism positions. To support vehicle-side integration, the system can transmit prism and tilt-compensated point of interest coordinates on the Controller Area Network bus, allowing the point of interest to be treated as a virtual position sensor (e.g., co-located with a rear-axle reference point). Experiments with a fixed ground reference point, using a prism to point of interest lever arm of approximately $1.07m$ and manual roll/pitch excursions of up to $60$°, yield three-dimensional root-mean-square errors between $2.9mm$ and $23.6mm$ across five test series. The results demonstrate that IMU-based tilt compensation enables reference measurements suitable for validating centimeter-level navigation systems under dynamic field conditions.

 \end{abstract}


\end{frontmatter}

\section{Introduction}\label{sec:introduction}

Accurate reference trajectories are required to validate autonomous agricultural robots and highly automated off-road vehicles under real-world field conditions. In this context, robotic total stations (RTS) are among the few sensors capable of providing continuous three-dimensional reference data with millimeter-level accuracy.

Although RTS observations provide highly accurate prism center coordinates, the relevant point of interest (POI) on the vehicle is typically located at a ground-level reference point, such as the origin of the vehicle coordinate system, the center of a tool track, or the tip of a measuring pole. This POI is commonly displaced from the prism center by a lever arm on the decimeter to meter scale. Since RTS measurements provide position information ($3D$) but not full pose information ($6D$), roll and pitch motions of the vehicle or mounting structure cannot be accounted for by the total station alone. As a result, even small tilts caused by uneven ground, driving dynamics, or human handling can introduce significant horizontal POI errors relative to the prism's vertically projected position. In practice, such uncompensated tilt effects can grow into the centimeter to decimeter range, despite the intrinsic measurement accuracy of the RTS.
\begin{figure}[!b]
  \centering
  \includegraphics[width=0.95\linewidth]{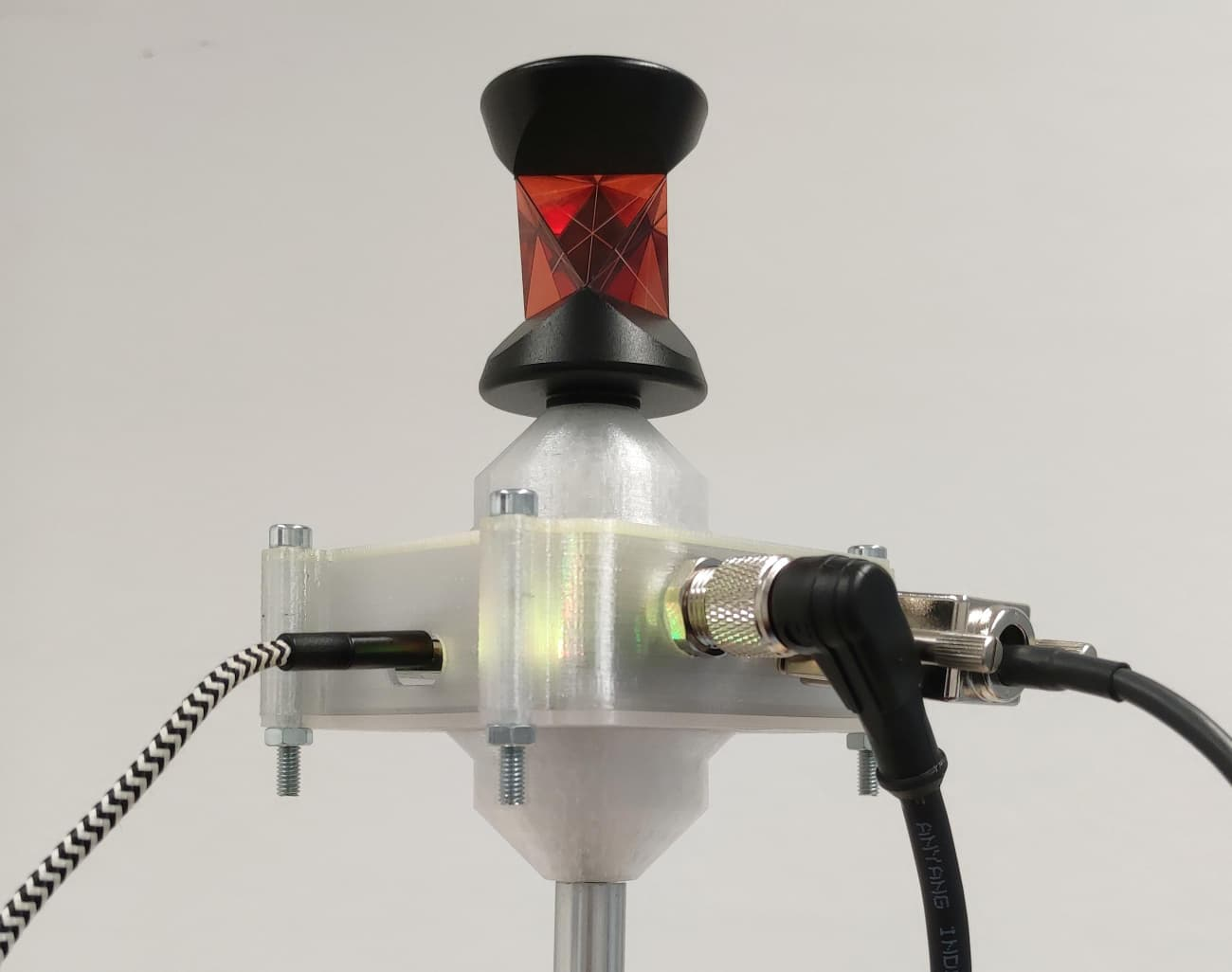}
  \caption{Smart Prism fully assembled with interface connections in operation.}
  \label{fig:prototype}
\end{figure}
Mechanical countermeasures, such as reduced driving speeds or additional stabilization structures, are only partially effective and often contradict realistic test conditions. The demand for highly accurate reference trajectories is particularly pronounced in dynamic scenarios such as lane changes, turning maneuvers, and operation on slopes, where conventional RTS prism setups reach their fundamental limits.

To address this limitation, the approach presented in this work combines prism center positioning with direct orientation acquisition. By integrating a MEMS-IMU into the prism, a Smart Prism system is created that provides the RTS prism position together with real-time roll and pitch angles. Using kinematic models and coordinate transformations, the tilt-corrected position of an arbitrarily definable reference point can be computed from this information, enabling reference measurements suitable for validating centimeter-level navigation systems under dynamic field conditions. In addition, the work is motivated by the practical integration of RTS-based reference measurements into vehicle electronic architectures. In mobile machinery, position sources are typically consumed as cyclic messages on the Controller Area Network (CAN) (e.g., from RTK-GNSS receivers), whereas RTS devices provide surveying-specific interfaces. Bridging this interface gap is therefore essential for integrating RTS-derived reference trajectories into vehicle-integrated validation setups.

The system presented in this whitepaper is based on a master's thesis \cite{Sharma.2025} that covers the development cycle from requirements analysis through hardware and firmware design to experimental validation. This paper focuses on the key technical contributions and experimental results, providing a practical solution for high-accuracy, tilt-compensated reference measurements in agricultural robotics.

\section{Hardware architecture}\label{sec:hardware}

\subsection{Computing Unit and Peripherals}\label{sec:hardware_mcu}
The central processing unit is an STM32H723ZG (ARM Cortex-M7, up to 550 MHz, with 1 MB flash and 564 KB RAM). The MCU provides a double-precision floating-point unit and hardware accelerators (CORDIC, FMAC) that support real-time trigonometric computations and filter operations required for attitude estimation and coordinate transformations.

The STM32H723ZG's peripherals allow for the direct connection of all relevant sensors and interfaces. The IMU is read via an SPI interface, the total station data is received from the Leica CS20 field controller via an RS-232 connection, and a CAN interface enables connection to the vehicle bus. A virtual COM port for the PC application is provided via USB-CDC.

\subsection{Inertial Sensors}\label{sec:hardware_imu}
A Murata SCH16T-K01 IMU (3-axis gyroscope and 3-axis accelerometer) is used to estimate roll and pitch. The selection was guided by expected bias instability, noise density, and temperature range, as these parameters directly affect drift and short-term noise in attitude estimation. According to the manufacturer specifications, the gyroscope bias instability is approximately $0.3\,\mathrm{^\circ/h}$ and the noise density is $0.0005\,\mathrm{^\circ/(s\sqrt{Hz})}$, with an operating temperature range of $-40\,\mathrm{^\circ C}$ to $110\,\mathrm{^\circ C}$. This performance is expected to be sufficient to stabilize roll and pitch estimation over the time scales relevant for RTS-based reference measurements.

\subsection{Power Supply and Protection Concept}\label{sec:hardware_power}
The system is designed for operation on a $12\,\mathrm{V}$ vehicle electrical system that meets the requirements of ISO 16750-2 \cite{ISO16750-2:2023} and can also be powered via a $5\,\mathrm{V}$ USB-C supply. ESD protection diodes (USBLC6-4, STMicroelectronics) and transient absorbers protect both the $12\,\mathrm{V}$ input and the $5\,\mathrm{V}$ USB line from overvoltage pulses.

An AP74502Q (Diodes Incorporated) protection controller monitors the input voltage, controls two back-to-back N-channel MOSFETs, and reliably disconnects the load from the vehicle electrical system in case of overvoltage, undervoltage, or reverse polarity. The design of the voltage dividers at the overvoltage and undervoltage inputs ensures that sustained voltages above approximately $16\,\mathrm{V}$ and voltages below approximately $4.5\,\mathrm{V}$ lead to the safe shutdown of the system.

The $3.3\,\mathrm{V}$ system voltage is generated via a low-dropout regulator (S-1214 series, ABLIC), which can be powered by either the vehicle's electrical system or the USB interface. Performance estimates and validation measurements show that the regulator operates at the upper end of its permissible thermal range under typical loads of approximately $0.9\,\mathrm{W}$ to $1.3\,\mathrm{W}$ of power dissipation. This necessitates targeted cooling of the regulator at high ambient temperatures or with an additional load.

\subsection{Interfaces and Mechanical Integration}\label{sec:hardware_interfaces}
For CAN communication, an SN65HVD230 (Texas Instruments) transceiver is used, operating in slope control mode to limit signal edges and reduce electromagnetic emissions. The physical layer conforms to ISO 11898-2 \cite{ISO11898-2:2024}, while the application layer commonly used in agriculture is defined by the ISO 11783 (ISOBUS) standard family \cite{ISO11783-1:2017}, which governs standardized data exchange between tractors, implements, and control systems based on CAN.

\begin{figure}
    \centering
    \includegraphics[width=1\linewidth]{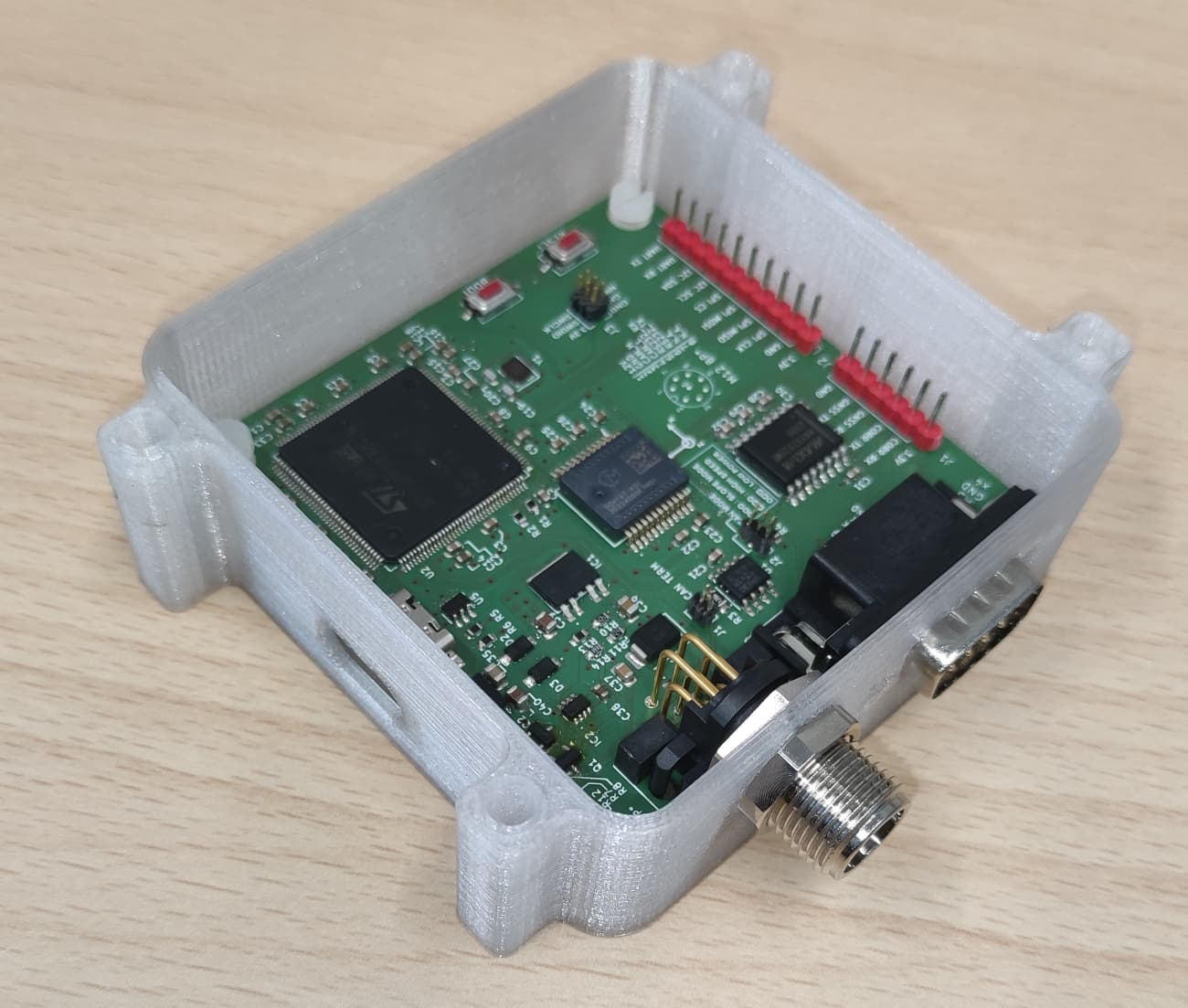}
    \caption{Smart Prism PCB, consisting of 90 individual components.}
    \label{fig:enter-label}
\end{figure}

Total station connectivity is provided via a MAX3232 (Analog Devices) RS-232 level converter, and a USB-C port establishes the PC connection. All hardware is integrated on a $70\,\mathrm{mm}\times 70\,\mathrm{mm}$ printed circuit board (PCB) using a double-sided FR4 design with a thickness of $1.6\,\mathrm{mm}$. The back side of the PCB largely forms a continuous ground plane to improve electromagnetic compatibility and signal integrity. 

The IMU is positioned close to the mechanical center of the housing to minimize kinematic offset effects. Current traces and thermal dissipation paths are dimensioned according to the estimated maximum load of approximately $0.4\,\mathrm{A}$. Critical components such as the voltage regulator are connected to the ground plane via enlarged copper areas and thermal vias to enhance heat dissipation.

For integration into a vehicle, an M12 8-pin screw connector is provided, which includes the $12\,\mathrm{V}$ power supply, connection to the vehicle's CAN bus, and an RS-232 interface as input for RTK correction signals for future GNSS integration. Firmware updates, $5\,\mathrm{V}$ operation, and debugging can be performed via a USB-C connection. The RS-232 signal from the CS20 device is connected via a DE9 connector.

The housing is additively manufactured from a UV-resistant material (PMMA) and comprises a main body to hold the PCB, a lid with an integrated prism mount, and a lower section that serves as the interface to the vehicle or tripod. The design allows for a secure mechanical connection of the prism, electronics, and mounting point without obstructing the view of the RTS. The housing is partially transparent, allowing light to shine through to three status LEDs on the PCB (see Figure \ref{fig:prototype}). The prism is mounted coaxially above the IMU to minimize lateral offsets and simplify the kinematic model. Figure \ref{fig:enter-label} shows the manufactured PCB. 

\section{Fusion Logic and Tilt Compensation}\label{sec:fusion}

\subsection{Adaptive Complementary Filter}\label{sec:acf}

The roll $\varphi_i$ and pitch $\theta_i$ angles are determined via an adaptive complementary filter \cite{Mahony.2008} through the fusion of gyroscope and accelerometer according to:

\begin{align}
\varphi_i = \alpha_{dyn} \cdot (\varphi_{i-1} + \dot{\varphi}_{gyro} \cdot \Delta t) + (1 - \alpha_{dyn}) \cdot \varphi_{acc}
\end{align}
\begin{align}
\theta_i = \alpha_{dyn} \cdot (\theta_{i-1} + \dot{\theta}_{gyro} \cdot \Delta t) + (1 - \alpha_{dyn}) \cdot \theta_{acc}
\end{align}

Here, $\dot{\varphi}_{gyro}$ and $\dot{\theta}_{gyro}$ denote the measured angular velocity in the respective axis and $\Delta t$ the sampling interval. The roll and pitch inclinations from the acceleration measurement are geometrically derived from the acceleration components in the $x, y, z$ directions of the IMU \cite{Munguía.2014}.

\begin{align}
\vec{a} = \begin{pmatrix} a_x \\ a_y \\ a_z \end{pmatrix}
\end{align}
\begin{align}
\varphi_{acc} = \operatorname{atan2}(a_y, a_z)
\end{align}
\begin{align}
\theta_{acc} = \operatorname{atan2}\left(-a_x, \sqrt{a_y^2 + a_z^2}\right)
\end{align}

In the adaptive complementary filter, a dynamic confidence value is used instead of a static value for the gyroscope. This value is determined based on the deviation of the norm of the measured acceleration vector from the local acceleration due to gravity, $g$.

\begin{align}
\Delta a = \left|  \parallel\vec{a}\parallel - g  \right|
\end{align}

The dynamic filter constant $\alpha_{dyn}$  is derived, which varies between a base value $\alpha_{base} = 0.9$ (steady-state case) and $1.0$ (strong dynamics). It is fixed at $1.0$ above a threshold of $\Delta a_{threshold} = 1 m/s^2 $  :

\begin{align}
\alpha_{dyn} = \min\left(1.0,  \alpha_{base} + (1-\alpha_{base}) \cdot \frac{\Delta a}{\Delta a_{threshold}}\right)
\end{align}

In the prototype, the yaw angle $\psi$ is determined solely by integrating the $z$-component of the gyroscope and is therefore subject to gradual drift. Automatic support via magnetometer or GNSS compass is not implemented. The yaw angle can be manually corrected via external commands if necessary. Although $\psi$ is logged by the system, it is not used for lever-arm rotation in the controlled tilt-compensation experiment; for this validation, $\psi$ is held constant (set to its initial value).

\subsection{Kinematics of lever arm transformation}\label{sec:kinematics}

\begin{figure}
    \centering
    \includegraphics[width=1\linewidth]{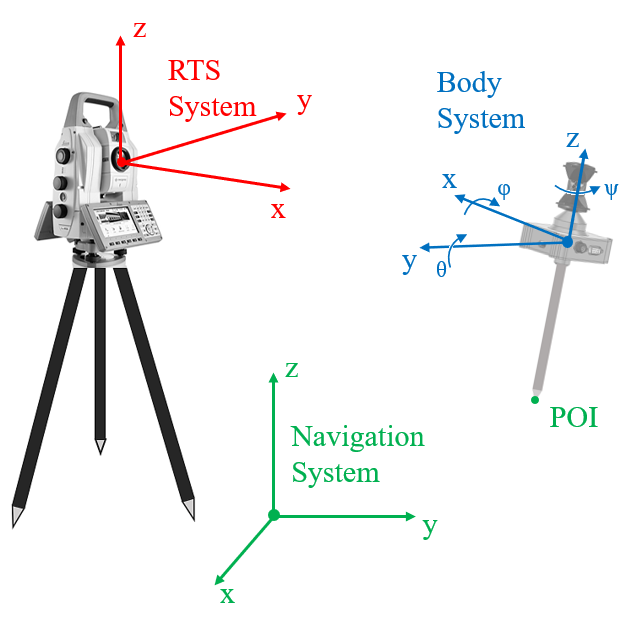}
    \caption{Conceptual visualization of the body, navigation and RTS system and POI}
    \label{fig:coordinate systems}
\end{figure}

The geometric basis (shown in Figure \ref{fig:coordinate systems}) of tilt compensation is the transformation of lever arms between a body-fixed coordinate system, which is tracked by the IMU and prism, and a stationary navigation system \cite{Durham.2013}. The orientation of the body relative to the navigation system is described by the Euler angles roll ${}^{N}\varphi$, pitch ${}^{N}\theta$, and yaw ${}^{N}\psi$. In the $Z-Y-X$ convention used here, the rotation matrix from the body-fixed system ($B$) to the navigation system ($N$) is the product of three axis rotations:

\begin{align}
\boldsymbol{R}_{B \to N} = \boldsymbol{R}_Z({}^{N}\psi) \cdot \boldsymbol{R}_Y({}^{N}\theta) \cdot \boldsymbol{R}_X({}^{N}\varphi)
\end{align}

The individual rotation matrices together form a complete orthogonal transformation that can convert any vector expressed in the body-fixed system into the navigation system.

In the Smart Prism system, three geometric points are considered: the center of the IMU, the center of the prism, and the POI. The rigid lever arms ${}^{B}L_{IMU \to Prism}$ and ${}^{B}L_{IMU \to POI}$ are in the body-fixed frame of reference. From these vectors, the lever arm from the prism center to the reference point in the body-fixed frame of reference is calculated:

\begin{align}
{}^{B}L_{Prism \to POI} = {}^{B}L_{IMU \to POI} - {}^{B}L_{IMU \to Prism}
\end{align}

The actual tilt compensation consists of transforming this into the navigation system using the current rotation matrix:

\begin{align}
{}^{N}L_{Prism \to POI} = \boldsymbol{R}_{B \to N} \cdot {}^{B}L_{Prism \to POI}
\end{align}

The resulting vector ${}^{N}L_{Prism \to POI}$ describes the spatial offset between the prism center and the reference point in the fixed coordinate system and is congruent with the current orientation of the system at all times.

\subsection{Fusion of RTS Data and IMU Orientation}\label{sec:fusion_rts_imu}

The fusion of RTS measurements with the IMU angles occurs in two steps: First, the polar measurements from the total station are converted into Cartesian coordinates, and then the oriented lever arm is added.

The RTS provides the slant distance $D$, the horizontal angle $Hz$, and the zenith angle $V$ for each measurement time. Cartesian prism coordinates in the total station system are calculated from these polar quantities. The horizontal distance $R$ and the height component $z$ are derived from the slant distance and the zenith angle as follows:

\begin{align}
R = D \cdot \sin(V)
\end{align}
\begin{align}
z = D \cdot \cos(V)
\end{align}

The horizontal components $x$ and $y$ follow from the horizontal distance and the horizontal angle.

\begin{align}
x = R \cdot \cos(Hz)
\end{align}
\begin{align}
y = R \cdot \sin(Hz)
\end{align}

To transfer the calculated prism coordinates into the higher-level navigation system, the three-dimensional Helmert transformation \cite{Chang.2016} is applied. It describes the relationship between the total station system and the navigation system using a scale factor $s$, a rotation matrix $\boldsymbol{R}$, and a translation vector $\boldsymbol{t}$. These parameters are determined in a calibration process from at least three spatially independent pairs of points using a closed adjustment solution. If ${}^{RTS}P_{Prism}$ is the prism position in the total station system, it is transformed into the navigation system according to:

\begin{align}
{}^{N}P_{Prism} = s \cdot \boldsymbol{R} \cdot {}^{RTS}P_{Prism} + \boldsymbol{t}
\end{align}

The tilt-corrected position of the reference point is ultimately obtained by adding the lever arm aligned in the fixed navigation system to the prism position:

\begin{align}
{}^{N}P_{POI} = {}^{N}P_{Prism} + {}^{N}L_{Prism \to POI}
\label{eq:poi}
\end{align}

Equation (\ref{eq:poi}) forms the core of the entire compensation procedure. It links the highly accurate position information from the RTS with the orientation information from the IMU to create a reference coordinate that reflects the actual location of the POI.

\section{Firmware Architecture and Signal Processing}\label{sec:firmware}

\subsection{Real-time Processing Firmware}\label{sec:rt_firmware}
The software architecture of the Smart Prism system is structured in two stages. An event-driven firmware runs on the STM32 microcontroller, reading sensor data, performing preprocessing and filtering, and managing the communication interfaces. A Python application on an external PC complements this with the kinematics of the tilt compensation, coordinate transformation, visualization, and data recording.

The STM32 firmware is implemented as a main loop with time-controlled task scheduling. Cyclically, IMU data is retrieved via SPI at a rate of approximately $100\,\mathrm{Hz}$, checksums and status bits are verified, and the raw data is converted into physical quantities. After data acquisition, the acceleration and gyroscope measurements pass through the adaptive complementary filter described in \cref{sec:acf}, ensuring that current roll and pitch angles are available at each sampling point.

Gyroscope bias calibration is performed at system startup by averaging $(N=1000)$ the gyroscope data over a defined idle period. The determined offsets are then subtracted from each measurement.

In parallel with IMU processing, total station measurements, provided by the Leica CS20 field controller via RS-232 in ASCII format, are read into a ring buffer using UART interrupts, decoded, and made available as structured data sets for fusion. The asynchronous nature of the total station data, with a typical output rate of $5\,\mathrm{Hz}$, is decoupled from the significantly higher IMU rate by the ring buffer concept. For the real-time fusion, the most recent orientation estimate is correlated with the incoming RTS coordinate packet. While this introduces a small phase lag in highly dynamic maneuvers, it remains negligible for the typical motion speeds of agricultural machinery during validation runs.

The firmware operates a virtual COM port via USB-CDC, allowing both measurement data to be sent to the PC and configuration commands to be received. A CAN driver also enables the output of status information to a vehicle network, meaning that the Smart Prism can, in principle, be integrated into a vehicle architecture without a PC.

\subsection{Python-based Fusion Logic and UI}\label{sec:python_fusion}
The actual tilt compensation, including the lever arm transformation and the Helmert transformation described in \cref{sec:kinematics,sec:fusion_rts_imu}, is performed in the prototype using a Python application on the PC. This application continuously receives the IMU angles and total station positions processed by the microcontroller via USB, calculates the tilt-corrected reference point coordinates in real time according to equation (\ref{eq:poi}), and displays the results in a graphical user interface. All intermediate values are synchronously logged in a CSV file to enable subsequent analysis in external tools.

Moving the fusion logic to the PC is a deliberate prototyping decision that allows for rapid iteration cycles during development. The available computing power and the existing hardware accelerators of the STM32H7 are, in principle, sufficient to perform the lever arm calculation, the rotation matrices, and the Helmert transformation directly on the embedded system. Such a relocation would make the Smart Prism a self-contained sensor that feeds tilt-compensated coordinates directly into the vehicle network via CAN.

\begin{figure}
    \centering
    \includegraphics[width=1\linewidth]{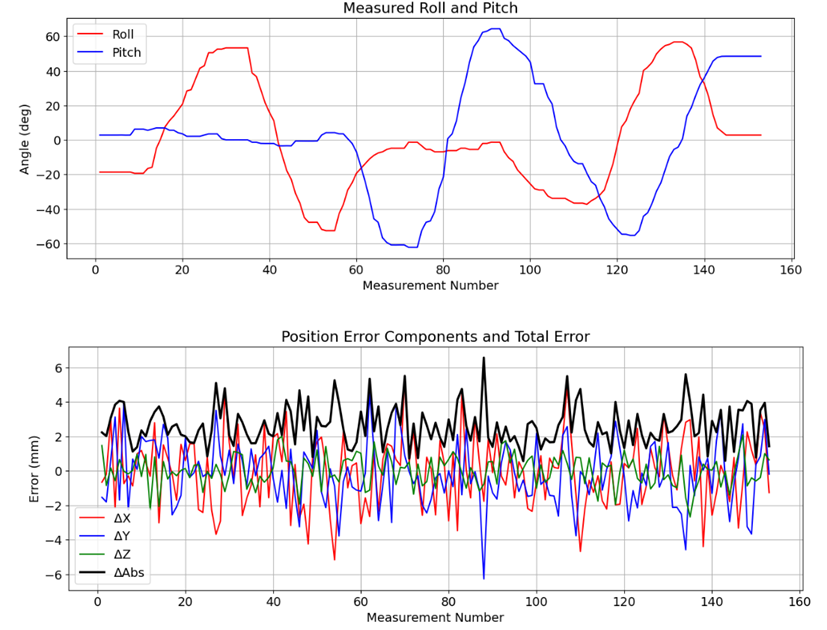}
    \caption{Roll/pitch trajectory calculation and associated positioning errors for a test ($ID=1$) with $n=154$ measurement points}
    \label{fig:plot}
\end{figure}

\section{Experimental Validation}\label{sec:validation}

\begin{table*}[!t]
\centering
\renewcommand{\arraystretch}{1.5}
\normalsize
\begin{tabularx}{\textwidth}{@{\extracolsep{\fill}} c ccc ccc ccc c c}
\hline
ID 
& $X$  & $Y$  & $Z$ 
& $\overline{\Delta X}$  & $\overline{\Delta Y}$  & $\overline{\Delta Z}$  
& $\sigma_{\Delta X}$ & $\sigma_{\Delta Y}$ & $\sigma_{\Delta Z}$  
& $\mathrm{\boldsymbol{RMSE}}_{3D}$ 
& $n$ \\
\hline

1 & 2131.2 & 998.3 &  -1537.4 
  & -0.046 & -0.160 &  0.006 
  &  2.037 &  1.809 & 0.906 
  &  2.867 & 154 \\

2 & 2423.8 & 778.5 & -1512.1 
  & -0.290 &  0.239 &  0.077 
  &  8.570 &  4.930 & 1.052 
  &  9.887 &  80 \\

3 & 2621.1 & 745.8 & -1520.4 
  & -0.337 & -1.072 & -1.651 
  &  7.967 &  4.779 & 9.652 
  & 13.472 &  92 \\

4 & 2152.2 & 954.5 & -1536.6 
  & -1.797 & -0.511 &  2.030 
  & 19.211 &  9.625 & 9.727 
  & 23.640 & 111 \\

5 & 1702.3 & 807.7 & -1521.2 
  & -2.184 & -0.422 &  0.483 
  & 10.566 &  9.644 & 9.767 
  & 17.406 & 134 \\

\hline
\end{tabularx} \newline 
\caption{Results of the tilt compensation test including reference coordinates. All coordinate and error quantities in this table are given in $mm$; $n$ denotes the number of samples.}
\label{tab:tilt_compensation}
\end{table*}

\subsection{Electrical and Interface Validation}\label{sec:validation_electrical}
The validation of the Smart Prism system pursues two complementary goals: First, it demonstrates that the electrical design of the power supply and communication interfaces operates stably under realistic load conditions. Second, it quantifies the extent to which the tilt compensation reduces the positional errors of the reference point caused by rod or mast tilts.

Laboratory tests were conducted on the power supply, in which the system was operated via the $12\,\mathrm{V}$ input with various loads and input voltages. The results confirmed that the protection controller reliably detects undervoltage, overvoltage, and reverse polarity, and disconnects the load when it falls outside the permissible operating range. Within the normal operating range of approximately $6\,\mathrm{V}$ to $16\,\mathrm{V}$, the power supply operates stably, although thermal reserves are limited under high loads.

The CAN interface was verified by exchanging test frames with PC-based CAN software. The RS-232 connection to the RTS via the CS20 controller was successfully tested as part of the overall integration, and the USB communication allowed for interference-free streaming of measurement data at the specified rates. This demonstrated the basic functionality of all communication paths.

\subsection{Experimental Setup for Tilt Compensation}\label{sec:validation_setup}
The central validation test investigates the system's ability to accurately reconstruct the reference point position despite large changes in tilt. For this purpose, the Smart Prism is mounted on a rod, the lower end of which forms the reference point (POI). 

The geometric offsets between the IMU, prism center, and POI are determined from CAD data. The distance between the IMU center and the prism center is approximately $75.6mm$. The offset from the IMU center to the POI at the rod base is approximately $992.0mm$. This yields a prism-to-POI lever arm magnitude of approximately $1067.6mm$.

The reference point is positioned at a fixed point on the ground and remains unchanged throughout the entire measurement series. First, the rod is aligned vertically to determine the reference coordinates in the total station system. Subsequently, the rod is continuously tilted manually in various directions and angles without laterally displacing or raising the base. For this controlled experiment, the RTS system is treated as the navigation system. Therefore, the Helmert transformation is set to identity ($s=1$, $\boldsymbol{R}=\boldsymbol{I}$, $\boldsymbol{t}=\mathbf{0}$). The RTS continuously records the prism center position at an output rate of $5\,\mathrm{Hz}$, while the IMU synchronously provides roll and pitch angles at $100\,\mathrm{Hz}$. The Python application continuously calculates the tilt-compensated reference point position from this data according to the fusion equation (\ref{eq:poi}) presented in \cref{sec:fusion_rts_imu}.

In total the experiment is repeated five times for different reference points, with $80$ to $154$ measurements being acquired per run. This setup replicates the intended application in an idealized form and allows for an isolated evaluation of the compensation performance without superimposed vehicle dynamics. Figure \ref{fig:plot} shows the measurement results as an example for one test run.

\subsection{Results of the Tilt Compensation}\label{sec:validation_results}
To assess accuracy, the differences between the calculated reference point position and the known reference coordinate are determined in the three spatial directions at all measurement times. The $3D$ root-mean-square error (${RMSE}_{3D}$) is used for the quantitative evaluation of compensation accuracy. It is defined as:

\begin{align}
{RMSE}_{3D} = \sqrt{\frac{1}{n}\sum_{i=1}^{n}\left(\Delta X_i^2 + \Delta Y_i^2 + \Delta Z_i^2\right)}
\end{align}

Here, $\Delta X$, $\Delta Y$, and $\Delta Z$ denote the deviations of the estimated reference point coordinates from the known reference coordinates in the three spatial directions. The ${RMSE}_{3D}$  serves as a summary measure of spatial positional accuracy and allows for a direct comparison of different test series.

Table~\ref{tab:tilt_compensation} summarizes the five test series. The obtained ${RMSE}_{3D}$ ranges from $2.9\,\mathrm{mm}$ to $23.6\,\mathrm{mm}$. Mean coordinate errors remain within a few millimeters in all runs, indicating no dominant systematic offset under the given calibration. The standard deviations vary between approximately $1\,\mathrm{mm}$ and $20\,\mathrm{mm}$, reflecting differences in excitation and handling during manual tilting.

These results confirm that the Smart Prism is capable of reconstructing a reference point position with errors in the lower centimeter range, even with considerable changes in tilt. The favorable mean values suggest a clean calibration of the lever arms and the absence of dominant systematic effects, while the variability of the ${RMSE}_{3D}$ values between the test series can be attributed to different handling conditions.

\subsection{Error Analysis and Influencing Factors}\label{sec:validation_error}
Analysis of the error characteristics shows that the residual deviations correlate strongly with manual test execution. In the test setup, the yaw angle is assumed to be constant, as only roll and pitch angles are stabilized by the complementary filter. In practice, however, small rotations of the rod around the vertical axis inevitably occur during manual handling, leading to systematic projection effects on the horizontal coordinates. These unmodeled yaw angle changes are identified as a significant source of error and, in particular, explain the higher ${RMSE}_{3D}$ values in individual test series where the rod was rotated more extensively.

In contrast, static IMU tests indicate very low noise in roll and pitch estimation: over $300s$ of steady-state measurement, the roll and pitch standard deviations remain on the order of a few thousandths of a degree. For a lever arm on the order of $1m$, this corresponds to sub-millimeter position noise. Therefore, the IMU sensor and the complementary filter are highly unlikely to be the dominant contributors to the observed position uncertainty.

Other potential sources of error include uncertainties in the lever arm length and temporal misalignment between IMU attitude estimates and RTS position packets due to asynchronous sampling and buffering. Such latency effects can lead to residual errors when the pole is tilted rapidly. An uncalibrated, generic prism was used in the experiment, which may introduce unknown centering errors. However, these are assumed to be negligible compared to the errors caused by manual execution and unobserved yaw dynamics.

Overall, the error analysis confirms that the quality of the tilt compensation itself is high and that the main areas for improvement lie in the yaw angle determination and the experimental methodology.

\section{Discussion}\label{sec:discussion}
The achieved ${RMSE}_{3D}$ values, ranging from approximately $3\,\mathrm{mm}$ to $24\,\mathrm{mm}$, must be evaluated within the context of the application. For the validation of RTK-GNSS systems, which themselves have a specified accuracy of one to two centimeters \cite{Vaidis.2021, Vieira.2022}, the demonstrated compensation accuracy is sufficient in many cases to serve as a reliable reference. For the highest accuracy requirements, such as the sub-centimeter evaluation of precision steering systems, improved yaw angle determination would be necessary to reduce the observed outliers.

\begin{figure}
    \centering
    \includegraphics[width=1\linewidth]{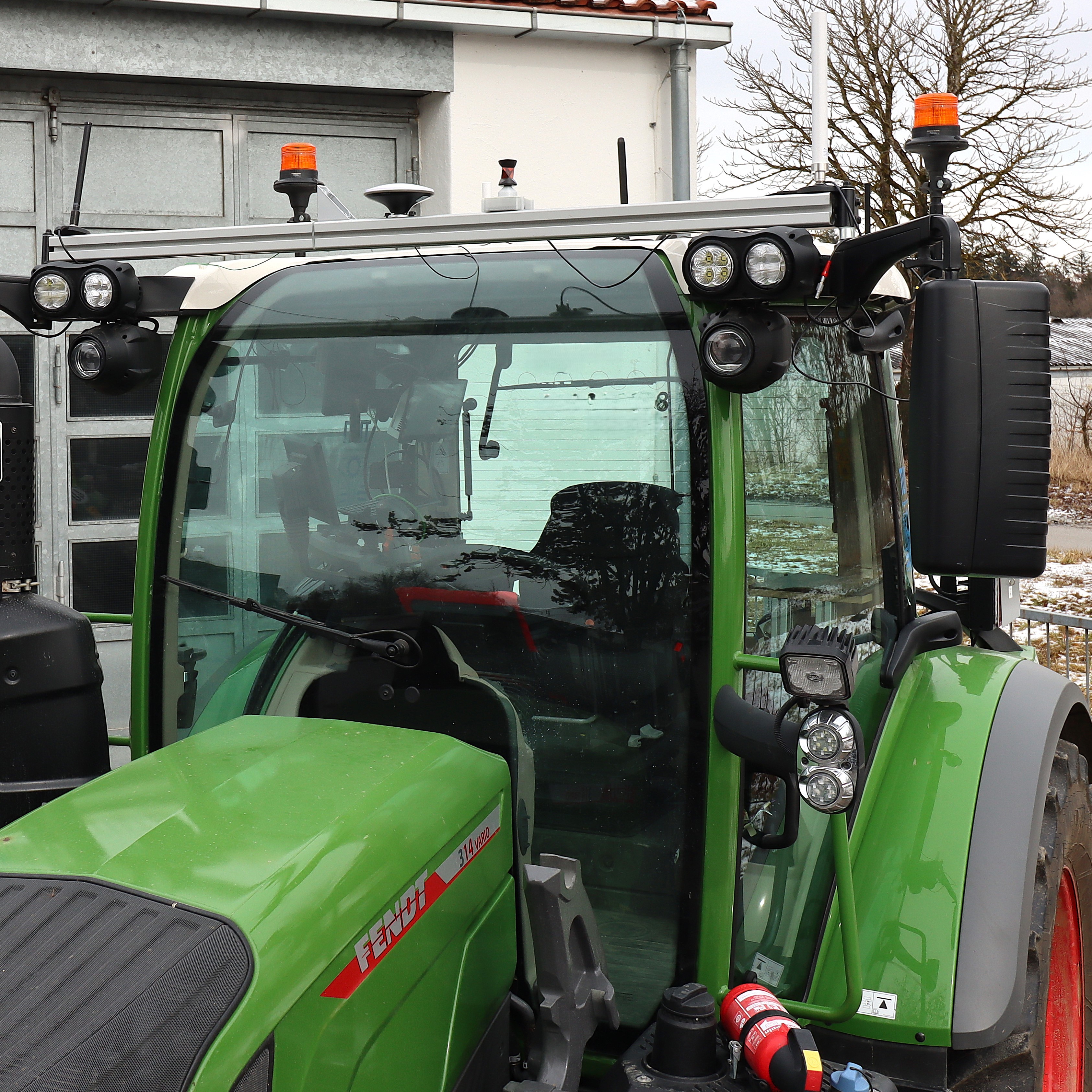}
    \caption{Concept of vehicle integration: The Smart Prism is mounted on the machine roof, co-located with a standard GNSS antenna.}
    \label{fig:tractor_integration}
\end{figure}

The yaw was treated as constant and was therefore not explicitly handled by the developed compensation pipeline. For deployment in vehicle-grade systems, a stable absolute heading input is required, typically provided by an additional external reference. Practical options include a 9-DOF IMU with magnetometer, GNSS-based heading using a dual-antenna setup \cite{Vieira.2022}, or short-term heading support derived from vehicle motion information such as steering angle and velocity available via the CAN bus. Figure \ref{fig:tractor_integration} illustrates a practical implementation scenario on a tractor, showing the Smart Prism co-located with a GNSS antenna, which facilitates such advanced fusion strategies

The two-stage architecture with microcontroller-based preprocessing and PC-based fusion logic represents a suitable solution for the prototype stage. The computing power of the STM32H7, with its double-precision FPU and trigonometric hardware accelerators, theoretically allows for a complete migration of the fusion algorithms to the embedded system. Such integration would not only eliminate the dependency on an external PC but also reduce the compensation latency and enable integration into real-time control loops in vehicles.

Comparison with existing solutions demonstrates the novelty of the approach. While commercial systems like the Leica AP20 AutoPole \cite{LeicaGeosystems.f} implement tilt compensation through internal sensors, they do not provide the open interfaces and integration possibilities required for agricultural research applications. Academic solutions \cite{Thalmann.2024, Vieira.2022} have demonstrated similar accuracy levels, but typically focus on specialized applications or require custom hardware platforms. The Smart Prism system bridges this gap by combining research-grade accuracy with practical vehicle integration through standard interfaces (CAN, ISOBUS) and a modular architecture suitable for agricultural machinery.

These findings confirm the validity of the overall approach and indicate that the remaining challenges are primarily engineering refinements rather than conceptual limitations.


\section{Conclusion}\label{sec:conclusion}
The Smart Prism system demonstrates that combining a RTS with a precise MEMS-IMU and a suitable fusion architecture provides a powerful foundation for tilt-compensated reference measurements in off-road applications. The developed prototype successfully addresses the fundamental limitation of conventional RTS setups: the positional errors caused by prism tilt that can reach centimeter to decimeter magnitudes depending on lever arm geometry.

Under idealized conditions with a fixed reference point and manual pole guidance, three-dimensional position errors in the range of approximately $3\,\mathrm{mm}$ to $24\,\mathrm{mm}$ are achieved, with mean coordinate errors consistently remaining in the low single-digit millimeter range. The achieved ${RMSE}_{3D}$ values align well with current studies in agricultural robotics, confirming that the fundamental approach is sound. The remaining deviations are primarily due to the lack of yaw angle support and do not represent a fundamental flaw in the compensation method.

The system builds upon existing surveying technology and can be implemented as a retrofit solution. The input stage, designed for vehicle electrical systems, the CAN interface for ISOBUS integration, and the modular software architecture enable expansion and adaptation to various application scenarios. The Smart Prism concept thus offers researchers and engineers in agricultural robotics a concrete approach to develop and validate autonomous agricultural and off-road systems.

The technical contributions of this work include:

\begin{enumerate}
\item Integrated Hardware Platform: A compact, vehicle-ready system combining high-performance microcontroller, precision IMU, automotive-grade power supply, and standard interfaces (CAN, RS-232, USB).
\item Real-Time Sensor Fusion: Implementation of adaptive complementary filtering with gyroscope bias compensation and dynamic adjustment based on acceleration magnitude.
\item Kinematic Tilt Compensation: Complete mathematical framework for transforming body-fixed lever arms to navigation-fixed coordinates and fusion with RTS polar measurements.
\item Experimental Validation: Quantitative assessment of tilt compensation accuracy under controlled conditions, demonstrating centimeter-level performance suitable for validating RTK-GNSS systems.
\end{enumerate}

Several areas for improvement have been identified, including robust yaw determination, migration of fusion algorithms to the microcontroller, GNSS integration, and validation on real vehicles. These areas form concrete starting points for future work.

Overall, the results support the proposed compensation concept and highlight yaw estimation as the main target for further improvement.

\begin{ack}
We would like to thank Michael Maier for his technical support, in particular for introducing the RTS setup and assisting with the manufacturing of the enclosure as well as the procurement of the required components.
\end{ack}

\bibliographystyle{IEEEtran} 
\bibliography{bibtex/literature}             

@inproceedings{Vaidis.2021,
  author    = {Vaidis, Maxime and Giguere, Philippe and Pomerleau, Francois and Kubelka, Vladimir},
  title     = {Accurate outdoor ground truth based on total stations},
  booktitle = {2021 18th Conference on Robots and Vision (CRV)},
  year      = {2021},
  pages     = {1--8},
  publisher = {IEEE},
  doi       = {10.1109/CRV52889.2021.00012},
  isbn      = {978-1-6654-1413-5}
}

@article{Mahony.2008,
  author  = {Mahony, Robert and Hamel, Tarek and Pflimlin, Jean-Michel},
  title   = {Nonlinear Complementary Filters on the Special Orthogonal Group},
  journal = {IEEE Transactions on Automatic Control},
  year    = {2008},
  volume  = {53},
  number  = {5},
  pages   = {1203--1218},
  issn    = {0018-9286},
  doi     = {10.1109/TAC.2008.923738}
}

@article{Thalmann.2024,
  author  = {Thalmann, Tomas and Neuner, Hans},
  title   = {Sensor fusion of robotic total station and inertial navigation system for 6DoF tracking applications},
  journal = {Applied Geomatics},
  year    = {2024},
  volume  = {16},
  number  = {4},
  pages   = {933--949},
  issn    = {1866-9298},
  doi     = {10.1007/s12518-024-00593-4}
}

@article{Vieira.2022,
  author  = {Vieira, David and Orjuela, Rodolfo and Spisser, Matthias and Basset, Michel},
  title   = {Positioning and Attitude determination for Precision Agriculture Robots based on IMU and Two RTK GPSs Sensor Fusion},
  journal = {IFAC-PapersOnLine},
  year    = {2022},
  volume  = {55},
  number  = {32},
  pages   = {60--65},
  issn    = {2405-8963},
  doi     = {10.1016/j.ifacol.2022.11.115}
}

@manual{LeicaGeosystems.f,
  author       = {{Leica Geosystems AG}},
  title        = {Leica AP20 Quick Guide},
  edition      = {1.0},
  year         = {2022},
  address      = {Heerbrugg, Schweiz},
  organization = {Leica Geosystems AG}
}

@article{Chang.2016,
  author  = {Chang, Guobin},
  title   = {Closed form least-squares solution to 3D symmetric Helmert transformation with rotational invariant covariance structure},
  journal = {Acta Geodaetica et Geophysica},
  year    = {2016},
  volume  = {51},
  number  = {2},
  pages   = {237--244},
  issn    = {2213-5812},
  doi     = {10.1007/s40328-015-0123-7}
}

@mastersthesis{Sharma.2025,
  author       = {Sharma, Sumesh},
  title        = {Development of a Smart Prism with IMU-Based Tilt Compensation for Accurate Positioning on Off-Road Vehicles using Robotic Total Stations},
  school       = {Technical University of Munich},
  address      = {Germany},
  year         = {2025},
  type         = {Master's Thesis},
  month        = dec,
  note         = {},
  language     = {english}
}

@book{Durham.2013,
 author = {Durham, Wayne},
 year = {2013},
 title = {Aircraft flight dynamics and control},
 address = {Chichester West Sussex},
 publisher = {{John Wiley {\&} Sons Inc}},
 isbn = {1118646800}
}

@article{Munguía.2014,
author = {Rodrigo Munguía and Antoni Grau},
title ={A Practical Method for Implementing an Attitude and Heading Reference System},

journal = {International Journal of Advanced Robotic Systems},
volume = {11},
number = {4},
pages = {62},
year = {2014},
doi = {10.5772/58463},

URL = { 
    
        https://doi.org/10.5772/58463
    
    

},
eprint = { 
    
        https://doi.org/10.5772/58463
    
    

}
}

@standard{ISO16750-2:2023,
  title        = {ISO 16750-2:2023 (E): Road vehicles --- Environmental conditions and testing for electrical and electronic equipment --- Part 2: Electrical loads},
  organization = {International Organization for Standardization},
  year         = {2023}
}

@standard{ISO11898-2:2024,
  title        = {Road vehicles --- Controller area network (CAN) --- Part 2: High-speed physical medium attachment (PMA) sublayer},
  organization = {International Organization for Standardization},
  year         = {2024}
}

@standard{ISO11783-1:2017,
  title        = {Tractors and machinery for agriculture and forestry --- Serial control and communications data network --- Part 1: General standard for mobile data communication},
  organization = {International Organization for Standardization},
  year         = {2017}
}

\end{document}